\documentclass[EPJ,twocolumn]{webofc}
\usepackage[varg]{txfonts}   
\usepackage{float}
\usepackage{fancyhdr}
\usepackage{mathrsfs}
\usepackage{comment}
\usepackage{hyperref}

\wocname {}
\woctitle{}
\title{Tau lepton reconstruction at the Muon Collider:
Cross section measurement of the $H\rightarrow\tau^+\tau^-$ process}

\author{Kevin Dewyspelaere\inst{1}\and Giacomo Da Molin\inst{1}\and Giovanni Battista Marozzo\inst{1}\and Michele Gallinaro\inst{1}}

\institute{
Laboratório de Instrumentação e Física Experimental de Partículas, Lisboa, Portugal
}

\abstract{Studies of Higgs boson properties are crucial for the understanding the Standard Model (SM), as it could couple to new particles and provide hints to physics Beyond the Standard Model (BSM). Different future colliders are proposed, among them the Muon Collider project would allow to perform unprecedented precision measurements of the Higgs boson parameters. The goal of this study is to estimate the statistical uncertainty of the cross section of the $H\rightarrow\tau^+\tau^-$ process at a 10 TeV center-of-mass energy Muon Collider. In order to reconstruct $\tau$ leptons in different decay modes,
the \texttt{TauFinder} algorithm is used. The efficiency of hadronic $\tau$ ($\tau_h$) lepton identification is estimated to be above 80\% for 1-prong and 50\% for 3-prong decay modes. This study focuses on the signal process $H\rightarrow\tau^+\tau^-$ in which the $\tau$ leptons decay hadronically ($\tau_h$). The main background processes are discussed and compared with the signal. The visible invariant mass is reconstructed and template fits are performed with Monte Carlo toy experiments. A statistical uncertainty on the cross section of the signal process: $\Delta\sigma/\sigma=$1.3\% is obtained. Finally, comparisons are made with the sensitivities at other future colliders, and possible improvements to the analysis are discussed.
}

\begin{document}

\twocolumn[
\begin{center}
{\Large\bfseries
Tau lepton reconstruction at the Muon Collider:\\
Cross section measurement of the $H\rightarrow\tau^+\tau^-$ process
\par}
\vspace{1em}

{\normalsize
Kevin Dewyspelaere,
Giacomo Da Molin,
Giovanni Battista Marozzo,
Michele Gallinaro
\par}

\vspace{0.5em}

{\itshape
Laboratório de Instrumentação e Física Experimental de Partículas, Lisboa, Portugal
\par}
\end{center}

\vspace{1em}

\begin{center}
\begin{minipage}{0.85\textwidth}
\small
\begin{abstract}

Studies of Higgs boson properties are crucial for the understanding of the Standard Model (SM), as the Higgs boson could couple to new particles and provide hints of physics beyond the Standard Model (BSM). Among the proposed future colliders, the Muon Collider would allow unprecedented precision measurements of Higgs boson parameters. The goal of this study is to estimate the statistical uncertainty on the cross section of the $H\rightarrow\tau^+\tau^-$ process at a center-of-mass energy of 10~TeV. Tau leptons are reconstructed using the \texttt{TauFinder} algorithm. The efficiency of hadronic $\tau$ ($\tau_h$) identification is found to be above 80\% for 1-prong and about 50\% for 3-prong decay modes. This study focuses on the fully hadronic final state of the $H\rightarrow\tau^+\tau^-$ decay. The main background processes are discussed and compared to the signal. The visible invariant mass is reconstructed and template fits are performed using Monte Carlo toy experiments. A relative statistical uncertainty on the signal cross section of $\Delta\sigma/\sigma = 1.3\%$ is obtained. Comparisons with sensitivities at other future colliders are also presented, and possible improvements to the analysis are discussed.

\end{abstract}
\end{minipage}
\end{center}

\vspace{2em}
]

\section{Introduction}
\label{sec:intro}
\subsection{The Higgs Physics and prospects for Future Colliders}
The Higgs boson (H), a key component of the Standard Model (SM), is a scalar neutral particle introduced by the Brout–Englert–Higgs mechanism ~\cite{broutengler}. It was first observed in 2012 by the ATLAS~\cite{higgsatlas} and CMS~\cite{higgscms} collaborations at the LHC. Its properties, including couplings, decay modes and mass, have been extensively studied, showing consistency with SM predictions, though precision is still limited for many channels. Key open questions remain, such as the measurement of Higgs self-couplings and the search for possible decays into beyond-the-Standard-Model (BSM) particles, including dark matter.  

To address these challenges, future collider projects, such as the HL-LHC ~\cite{hllhc}, FCC ~\cite{fcc}, CEPC ~\cite{cepc}, ILC ~\cite{ilc}, CLIC ~\cite{clic}, and Muon Collider~\cite{Muoncollider}, aim at producing Higgs bosons in unprecedented numbers and explore the properties with higher precision. These facilities will significantly improve sensitivity to couplings, in particular to $\tau$ leptons, and will probe Higgs self-interactions, offering unique opportunities to uncover possible deviations from the SM and explore the possible presence of new physics processes.

\subsection{The Muon Collider Project}
The Muon Collider (MuCol)~\cite{Muoncollider} represents a unique proposal among future colliders, as it would collide muons and antimuons rather than protons or electrons. Thanks to the large muon mass, synchrotron radiation is strongly suppressed, allowing multi-TeV collisions in a relatively compact ring, while retaining the clean leptonic environment typical of $e^+e^-$ machines. This makes the MuCol simultaneously a precision tool and a high-energy frontier collider. Its physics potential is particularly strong in the Higgs sector. At multi-TeV energies, Higgs boson production is dominated by vector boson fusion and Higgs-strahlung, enabling precise measurements of Higgs couplings and self-interactions with large statistics. 

\begin{figure}[H]
\centering
\sidecaption
\includegraphics[width=0.50\textwidth,clip]{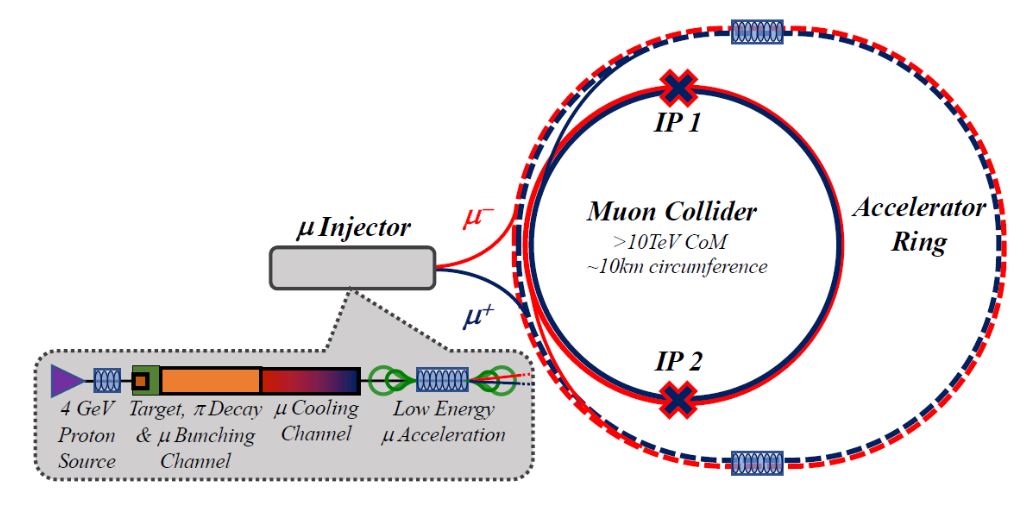}
\caption{Conceptual scheme of the MuCol facility as planned by the International Muon Collider Collaboration (IMCC). From Ref.~\cite{towardMuoncollider}.}
\label{fig:mucollfacility}       
\end{figure}

Figure~\ref{fig:mucollfacility} illustrates the conceptual scheme of the facility as designed by the International Muon Collider Collaboration (IMCC), which foresees staged operation at 3 TeV and 10 TeV, with luminosities sufficient to produce hundreds of thousands of Higgs bosons per year. Despite its advantages, the MuCol also faces major technological challenges, including production, cooling, and rapid acceleration of short-lived muon beams, as well as the mitigation of beam-induced backgrounds (BIB), originating from muon decays along the beamline that produce a large flux of secondary particles entering the detector. Nevertheless, its capability to combine energy reach with precision measurements makes it one of the most promising projects for exploring the Higgs sector and possible physics beyond the SM.

\section{Simulation and event reconstruction}
\subsection{The MAIA detector apparatus}
\begin{figure}[H]
\centering
\sidecaption
\includegraphics[width=0.48\textwidth,clip]{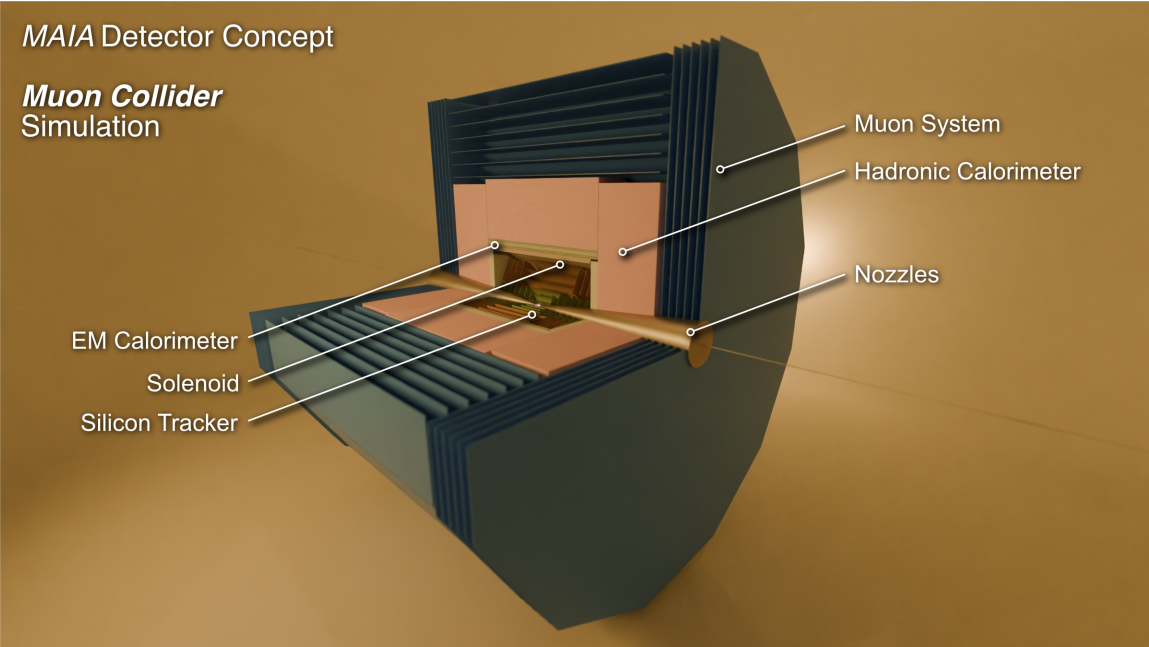}
\caption{Illustration of the MAIA detector layout. The detector is shown with a $\pi$/2 cutaway in $\phi$ for illustration (from Ref~\cite{maia}).}
\label{fig:maia}       
\end{figure}

The MAIA (Muon Accelerator Instrumented Apparatus) detector ~\cite{maia} is a new concept specifically designed for $ \sqrt{s} =$  10 TeV $\mu^+\mu^-$ collisions, aiming to cope with the unique challenges of a high-energy muon collider. Its layout combines an all-silicon tracker embedded in a 5~T solenoidal field, surrounded by high-granularity calorimeters: a silicon–tungsten electromagnetic calorimeter and an iron–scintillator hadronic calorimeter, optimized for particle-flow reconstruction. The outermost layer consists of an air-gap muon spectrometer, enabling precise standalone tracking of high-momentum muons (Fig.~\ref{fig:maia}).

A key feature of the apparatus is its capability to operate under intense BIB from muon decays.
Overall, MAIA is conceived as a general-purpose detector for the Muon Collider, providing both the hermeticity and precision required for Higgs and SM measurements, as well as sensitivity to new physics phenomena~\cite{newphysics}.

\subsection{Event reconstruction}

The reconstruction of physics objects at the Muon Collider is performed with a dedicated simulation, which proceeds from event generation to the creation of particle flow objects (PFOs) ready for analysis. Signal events are generated with external event generator tools such as MadGraph5~\cite{madgraph} with Pythia8 that is used for $\tau$ decay, hadronization and showering, and then passed through a full detector simulation based on GEANT4~\cite{geant4} within the ILCSoftware~\cite{ilcsoftware} framework adapted for MuCol~\cite{gitmucol}. Digitization modules implemented in MARLIN~\cite{marlin} convert simulated energy deposits into detector hits, applying realistic spatial and timing resolutions. In this study, BIB is not considered and only single interactions are included.

The calorimetric information is processed with the Pandora Particle Flow Algorithm (PandoraPFA)~\cite{pandorapfo}, which combines tracker and calorimeter data to reconstruct individual particles with optimal precision. The individual particles, i.e. Particle Flow Objects (PFOs), are then produced by associating tracks and clusters, assigning particle identities and preparing the reconstructed objects for physics analyses.

This reconstruction chain provides a realistic modeling of the detector response,
except for the beam induced background (BIB) which is not included in this study.

\section{Tau reconstruction and identification}
\label{sec:tau_reco_id}

\subsection{Tau lepton in the Standard Model}

The $\tau$ lepton is a third-generation fermion, and in the SM $\tau$ pairs can be produced in the decays of electroweak bosons, such as $H\rightarrow\tau^+\tau^-$, $Z\rightarrow\tau^+\tau^-$.

\begin{figure}[H]
\centering
\sidecaption
\includegraphics[width=0.48\textwidth,clip]{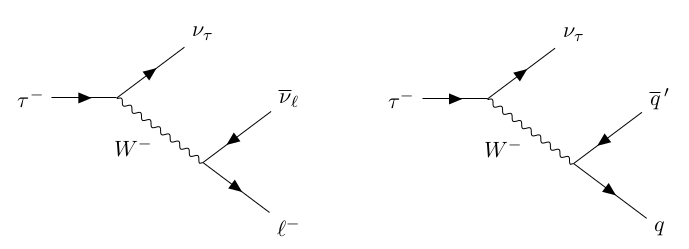}
\caption{Feynman diagrams for a $\tau^-$ lepton decaying into leptonic (left) and hadronic
(right) final states.}
\label{fig:taudecay}
\end{figure}

The $\tau$ lepton is unstable, with a short lifetime of about $2.9 \times 10^{-13}$~s~\cite{lifetau}, and decays via the weak interaction into a $\nu_\tau$ and a virtual W boson. The subsequent W decays either to a purely leptonic final state ($\tau_\ell$) or to a hadronic final state ($\tau_h$), as illustrated in Fig.~\ref{fig:taudecay}. 

\begin{table}[h!]
    \centering
    \includegraphics[width=\linewidth]{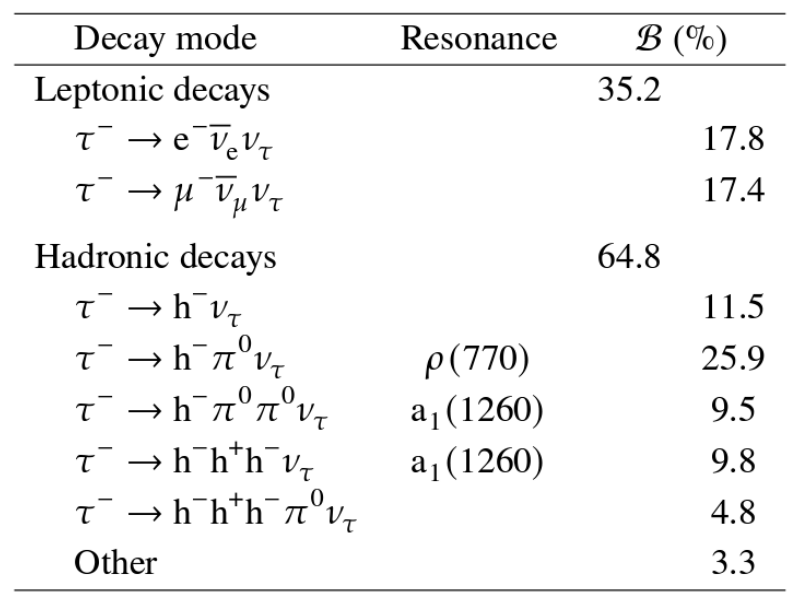} 
    \caption{
    Decays of $\tau$ leptons and their branching fractions in percentage \cite{decayBRtau} For
the decay modes with intermediate resonances, these are shown in the middle column.
Charged hadrons are denoted by the symbol $h^\pm$. Only $\tau^-$ decays are shown, since the decays and values of the branching fractions are identical for charge-conjugate decays.}
    \label{tab:table_taubr}
\end{table}

In the hadronic final state, the visible decay products are mostly light mesons ($\pi^\pm$, $K^\pm$, $\pi^0$), which result in one or three charged hadrons in the final state. This motivates the classification 
into \textbf{1-prong} and \textbf{3-prongs} categories.

The main SM decay modes of the $\tau$ lepton, together with their branching fractions, are summarized in Table~\ref{tab:table_taubr}. Roughly 35\% of $\tau$ decays are purely leptonic, while the remaining 65\% are hadronic, making hadronic $\tau$ reconstruction a key experimental challenge.

\subsection{The \texttt{TauFinder} algorithm}

\texttt{TauFinder}~\cite{taufinder} was originally developed for $\tau$ reconstruction in the CLIC experiment and is therefore optimized for lepton collider environments. It follows a cone-based jet-finding approach, using the four-momenta of charged and neutral reconstructed particles. The algorithm starts by selecting high-energy charged particles as seeds and builds a $\tau$ candidate by iteratively adding other charged and neutral particles within a narrow ``signal" cone 
of radius $\Delta R = 0.10$
around the seed direction, dynamically updating the cone axis. 
The cone size is defined as $\Delta R = \sqrt{(\Delta\eta)^2 + (\Delta\phi)^2}$,
where $\Delta\eta$ and $\Delta\phi$ are the differences of pseudo-rapidities and azimuthal angles.
Once all candidates are reconstructed, a merging step ensures that overlapping objects are combined.

To reduce misidentification, \texttt{TauFinder} applies a few quality cuts: the number of charged tracks must be either one or three (reflecting 1-prong and 3-prongs $\tau$ decays), the total number of particles from $\tau$ decays must be below ten, and the reconstructed candidate charge must be $\pm 1$. Additional reconstruction and isolation criteria are used to further suppress background: seeds must have transverse momentum $p_T > 5$~GeV, all associated particles $p_T > 1$~GeV, and an "isolation" cone ($0.10 < \Delta R < 0.40$) is built around the $\tau$ candidate. 

Candidates passing all requirements are identified as reconstructed $\tau$ leptons, with the charge distinguishing $\tau^+$ from $\tau^-$.
A study of the $\tau$ isolation is presented in Sec.~\ref{sec:taufakes}. No isolation requirement is applied in this efficiency study.

\subsection{\texttt{TauFinder} algorithm performance}

The performance of the \texttt{TauFinder} algorithm was evaluated using a sample of 15~000~$\tau$ ``particle-gun" ($\tau$-gun) events, equally split between $\tau^+$ and $\tau^-$. The generated $\tau$s were assigned kinematical properties uniformly distributed in $p_T \in [20, 320]$ GeV, $\phi \in [0, 2\pi]$, $\theta \in [10^\circ, 170^\circ]$. 
In the ``particle-gun" sample, 
the $\tau$ decay
is handled directly by GEANT4 during the detector simulation, after the generation step.

\texttt{TauFinder} does not distinguish between hadronic ($\tau_h$) and leptonic ($\tau_\ell$) decays at the reconstruction level; a basic classification can be introduced by requiring the presence of charged hadrons among the decay products. In addition, the current simulation framework does not reconstruct neutral pions explicitly: $\tau_h \rightarrow \pi^\pm + \pi^0$ decays are thus treated through their photon products. The classification of reconstructed $\tau_h$ was therefore performed only by prong multiplicity, i.e. counting the number of charged pions associated with each candidate.

In this study, only hadronic $\tau$ decays were selected to evaluate reconstruction efficiency. 

\begin{figure}[H]
\centering
\sidecaption
\includegraphics[width=0.48\textwidth,clip]{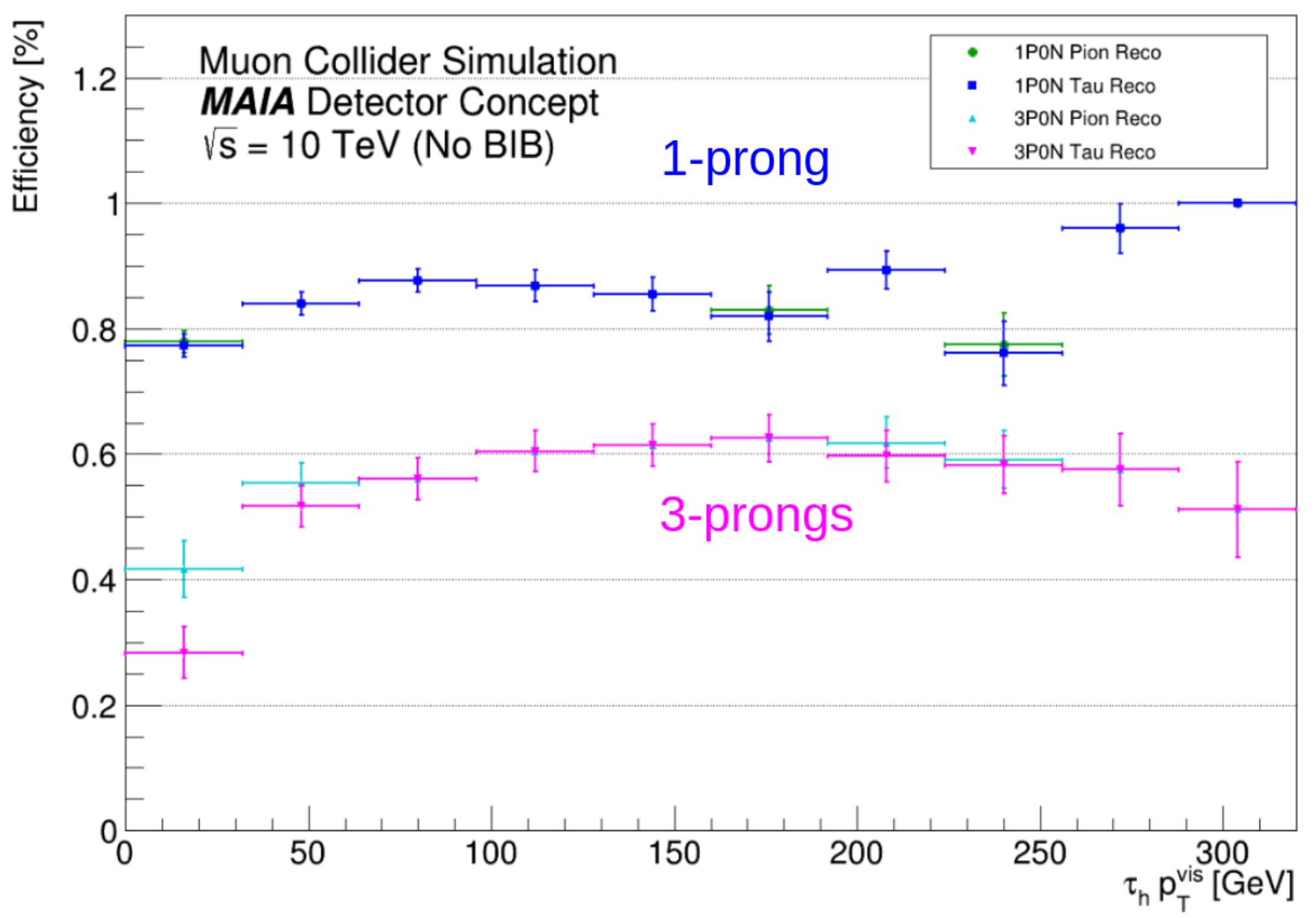}
\caption{Reconstruction efficiency of 1-prong and 3-prong $\tau$ decays with the \texttt{TauFinder} algorithm, as a function of the generated visible transverse momentum of the $\tau$ lepton. The efficiencies are compared to those obtained from charged pion reconstruction alone. Results are shown for Muon Collider simulations with the MAIA detector concept at $\sqrt{s} = 10$ TeV, without BIB.}
\label{fig:efficiency}       
\end{figure}

Figure~\ref{fig:efficiency} shows the resulting efficiency as a function of the generated visible transverse momentum $p_T^{vis}$, for both 1-prong and 3-prong $\tau$ decays. The results indicate that 1-prong $\tau$ decays are reconstructed with an efficiency close to 80–90\% over most of the $p_T^{vis}$ range, only slightly below the pion reconstruction baseline due to the additional cuts imposed by \texttt{TauFinder}. In contrast, 3-prong $\tau$ decays exhibit a significantly lower efficiency, around 50–60\%, with a mild dependence on $p_T^{vis}$. This reduction reflects the intrinsic challenge of reconstructing three charged prongs simultaneously, given detector acceptance and the isolation requirements of the algorithm. Overall, these results confirm that \texttt{TauFinder} provides a robust reconstruction of 1-prong $\tau_h$ candidates, while highlighting the need for further optimization to improve the efficiency for 3-prong modes.

\subsection{Electromagnetic Fraction (EMF)}
\label{sec:emf}

\begin{figure}[ht]
\centering
\includegraphics[width=0.48\textwidth]{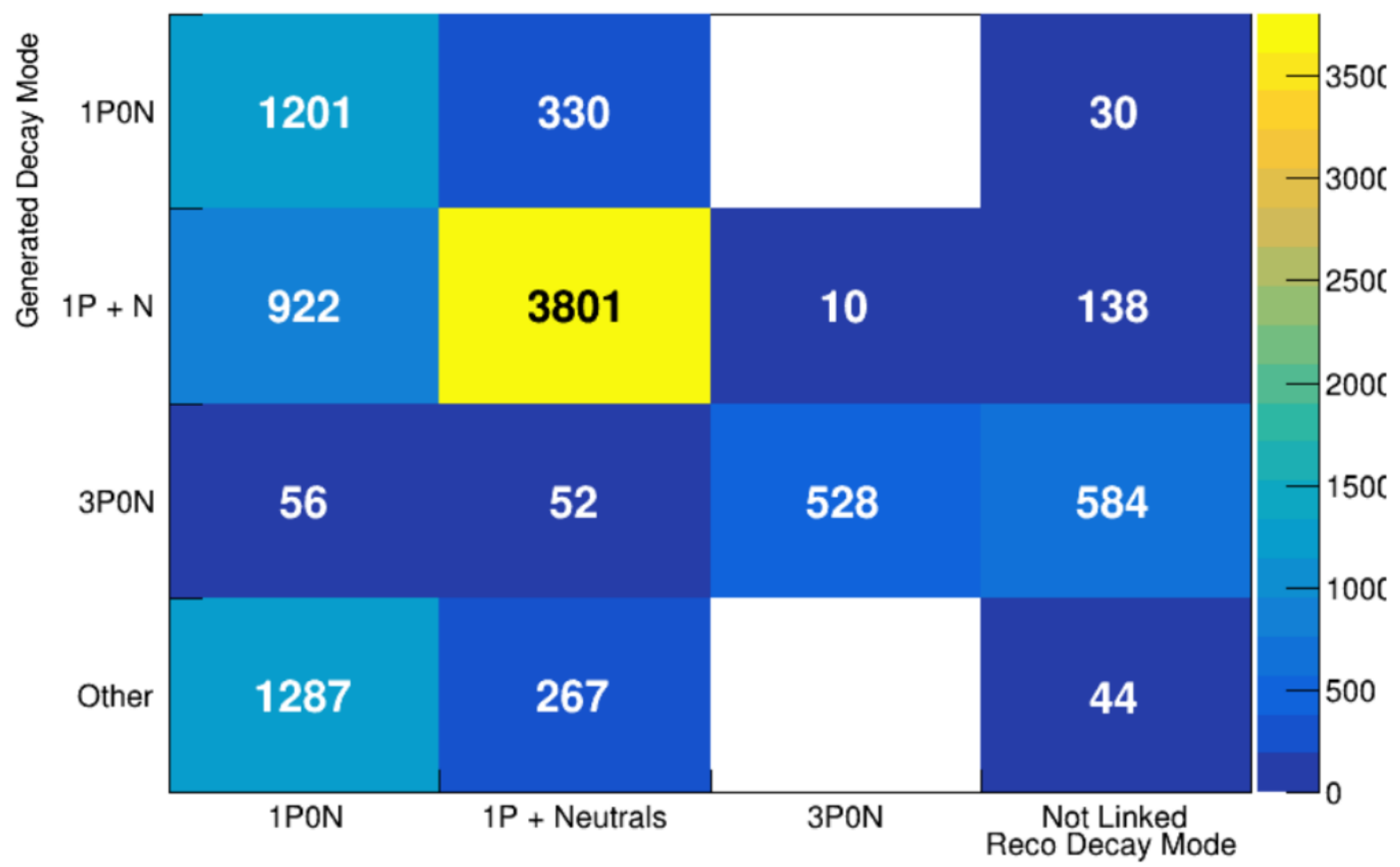}
\caption{Decay mode matrix before the EMF cut. A significant contamination of generator-level electron decays (included in the "other" category in the y-axis) is reconstructed as 1-prong hadronic $\tau$ decays, either as 1P0N (no neutrals) or 1P+N (with neutrals); after the EMF $<1$ cut, yields are largely suppressed (Fig.~\ref{fig:dmatrix_after}).}
\label{fig:dmatrix_before}
\end{figure}

It was observed that many $\tau$ leptons in the electron decay mode ($\tau\rightarrow e\nu_e\nu_\tau$) at
generator level were reconstructed as one-prong hadronic $\tau$ candidates. This effect can be
seen in Fig.~\ref{fig:dmatrix_before}, where the decay mode matrix shows a significant
population in the ``other'' category at generator level but is reconstructed as 1-prong $\tau_h$
either as 1P0N (no neutrals) or 1P+N (with neutrals).

\begin{figure}[ht]
\centering
\includegraphics[width=0.48\textwidth]{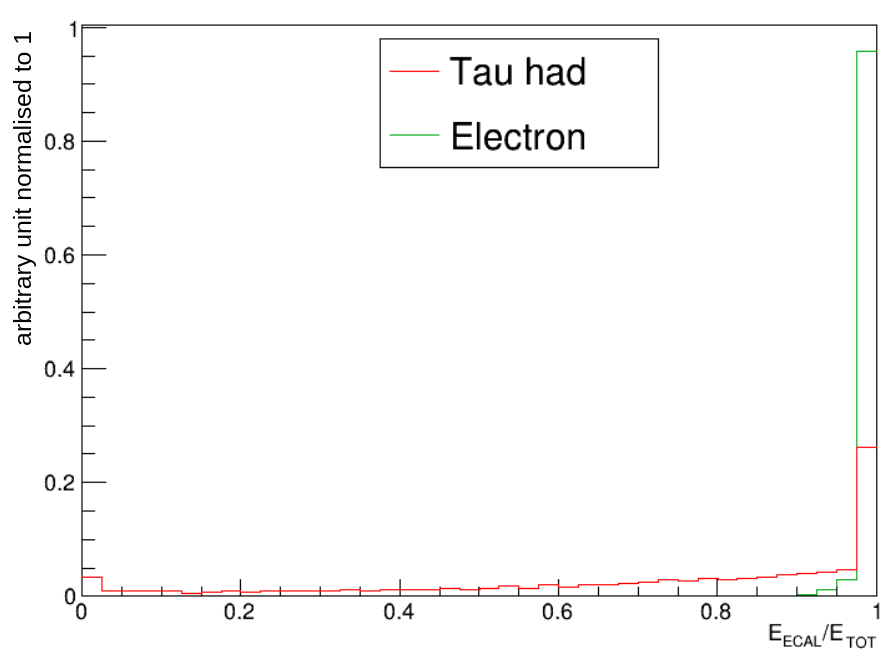}
\caption{Normalized EMF distributions for reconstructed $\tau$ candidates decaying hadronically (red) and to electrons (green). Electrons peak at $\mathrm{EMF}=1$, while $\sim 20\%$ of hadronically decaying $\tau$s deposit their full energy in the ECAL.}
\label{fig:emf_dist}
\end{figure}

To understand this misclassification, the electromagnetic fraction (EMF) associated to the $\tau_h$ energy cluster deposited in the ECAL and HCAL calorimeters was studied:
\begin{equation}
\mathrm{EMF} = \frac{E_{\mathrm{ECAL}}}{E_{\mathrm{ECAL}} + E_{\mathrm{HCAL}}}.
\end{equation}
The normalized EMF distributions for reconstructed hadronic $\tau$ candidates and electrons
are shown in Fig.~\ref{fig:emf_dist}. As expected, electrons peak at $\mathrm{EMF}=1$, since
they deposit most of their energy in the ECAL. About 20\% of reconstructed
hadronic $\tau$ candidates also appear at $\mathrm{EMF}=1$; a large fraction of these are
$\tau$ leptons in the electron decay mode misidentified as one-prong hadronic $\tau$'s.

\begin{figure}[ht]
\centering
\includegraphics[width=0.48\textwidth]{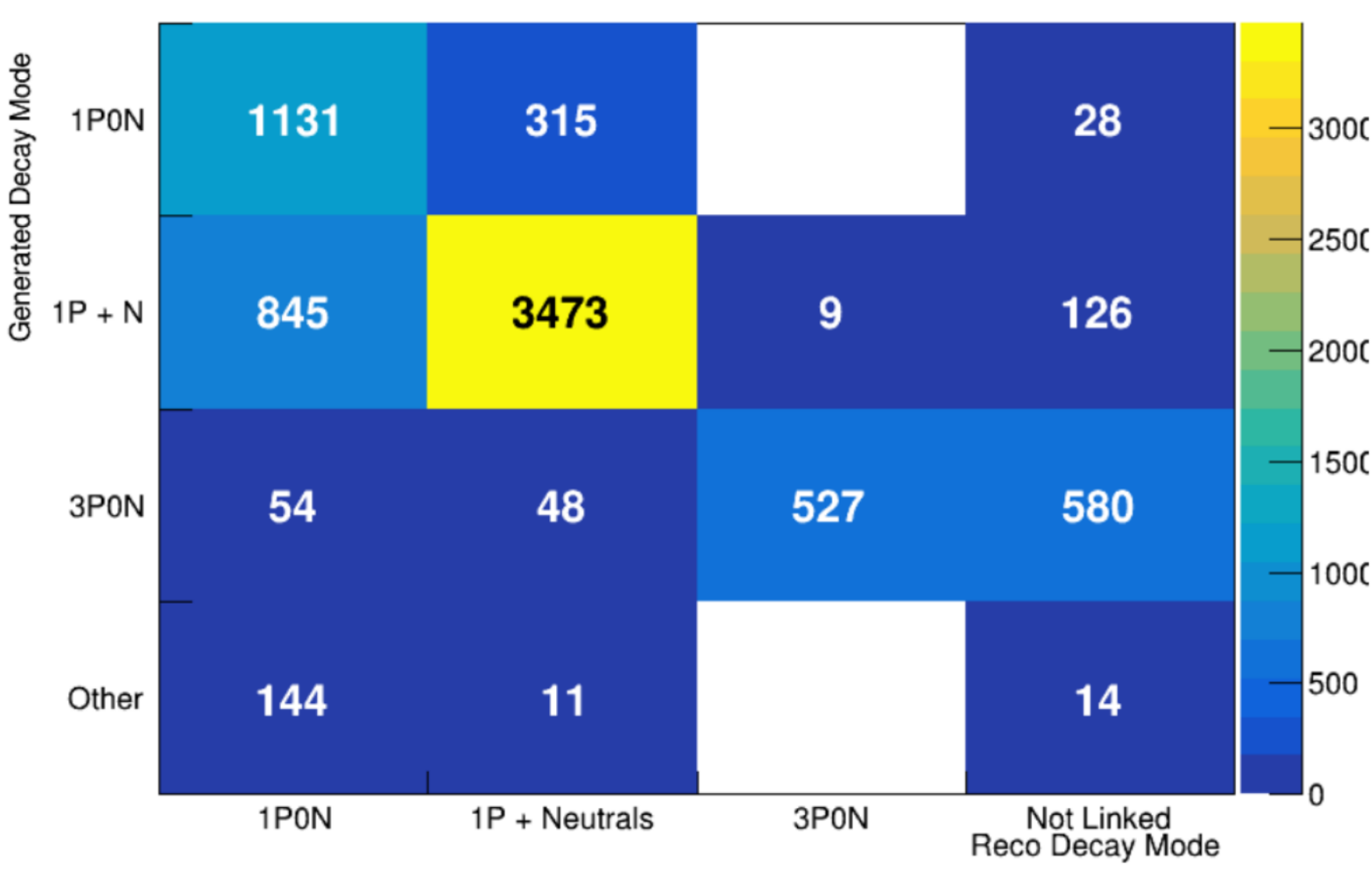}
\caption{Decay mode matrix after the EMF cut $\mathrm{EMF}<1.0$. The contamination from
generator-level electrons reconstructed as hadronic $\tau$'s is largely removed.}
\label{fig:dmatrix_after}
\end{figure}

To mitigate this effect, an EMF requirement of $\mathrm{EMF} < 1.0$ was applied. As shown in
Fig.~\ref{fig:dmatrix_after}, this cut removes the vast majority of events from the
``other'' category at generator level reconstructed as $1P0N$ or $1P+N$, thus significantly
improving the purity of the hadronic $\tau$ selection while keeping a high efficiency for
genuine hadronic $\tau$ decays.
This cut is applied as a simple patch to remove misidentified electrons and further improvements in the pion and electron ID are certainly necessary.

\subsection{Tau misidentification}
\label{sec:taufakes}

In the context of $\tau$ lepton reconstruction and identification, an important aspect is the misidentification rate, i.e. the probability that a physics object that is not a $\tau$ lepton (for instance, a genuine hadronic jet or an electron) is incorrectly reconstructed as a $\tau$ candidate. Such objects, also referred to as fake $\tau$, represent a potential source of background contamination in precision measurements such as the $H\rightarrow\tau^+\tau^-$ decay.
In this study, particular attention is given to the misidentification of hadronic $\tau$ candidates reconstructed by \texttt{TauFinder}.

\begin{figure}[ht]
\centering
\includegraphics[width=0.48\textwidth]{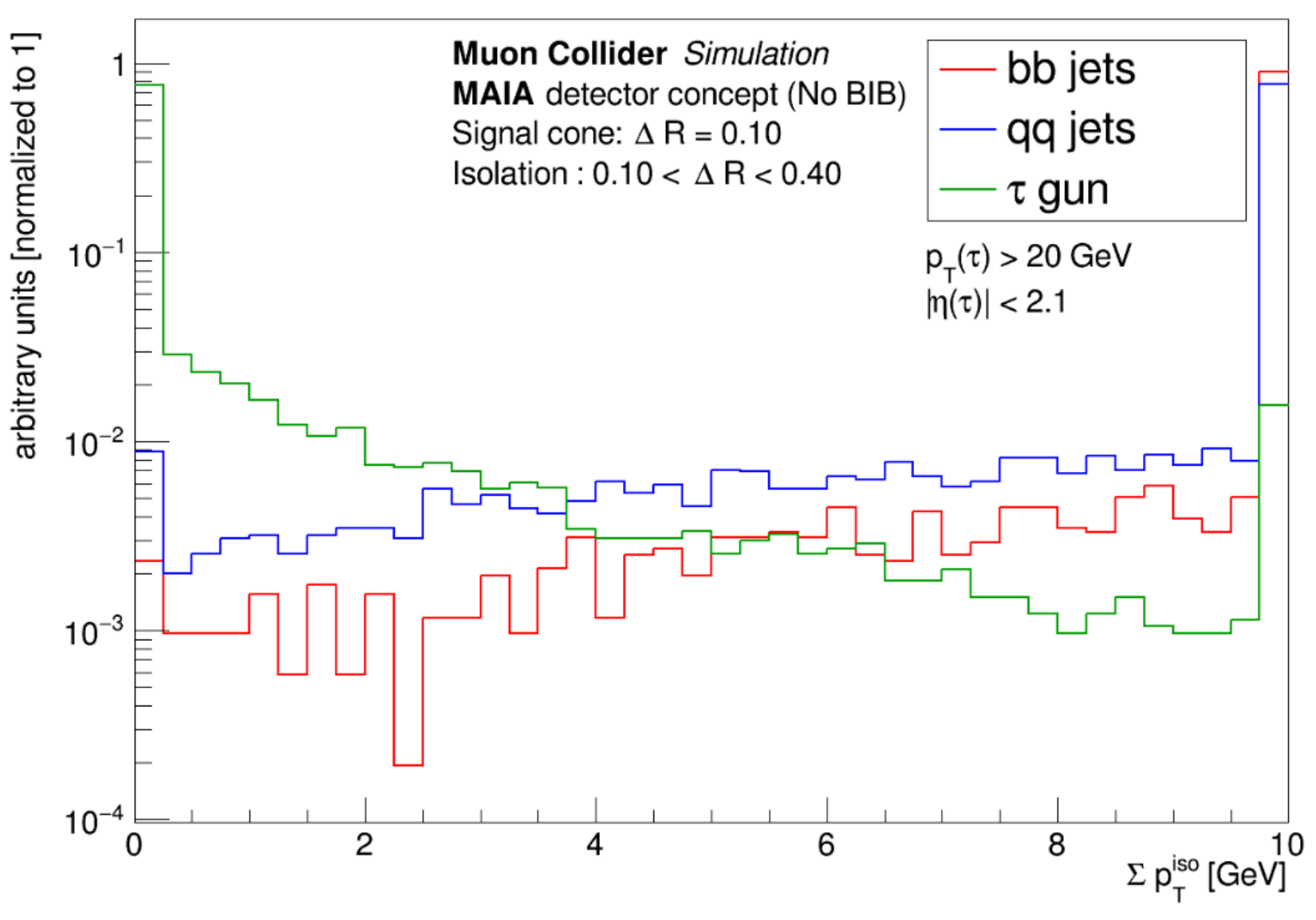}
\caption{Distribution of the energy deposited in the isolation cone around the reconstructed $\tau_h$ candidates for different physics object samples. The isolation energy is a key variable used to suppress misidentified $\tau$ originating from hadronic jets, as genuine $\tau$ typically exhibit lower energy activity in the surrounding cone.}
\label{fig:isoE}
\end{figure}

To quantify the fake $\tau_h$ rates, dedicated event samples were generated with MadGraph5: 
$Z\rightarrow q\bar{q}$ (light-flavor jets, 15k events), $Z\rightarrow b\bar{b}$ (heavy flavor, 15k events), $H\rightarrow \tau\bar{\tau}$ (signal). 
In each case, the generated jets were required to satisfy basic kinematical requirements such as: $p_T >20$~GeV, $|\eta| < 2.1$, $m_{inv}< 3$~GeV, number of reconstructed particle composing the $\tau_h$ candidate to be less than 8, EMF $< 1.0$. Additional requirements are applied, based on the energy inside the isolation cone (see Fig.~\ref{fig:isoE}), defined as the scalar sum of the tracks contained in the isolation cone ($0.10<\Delta R<0.40$) around the $\tau_h$ leading track.

The misidentification rate is defined as the ratio between the number of reconstructed $\tau_h$ candidates and the total number of generated jet-like objects that have $p_T >20$~GeV and $|\eta|<2.1$. The $\tau_h$ efficiency is determined by calculating the ratio of reconstructed divided by the generated $\tau_h$ objects.

\begin{table}[h!]
\centering
\begin{tabular}{cccc}
\hline\hline
 Physical objects & (u, s, d) jets & $b$-jets & $\tau_h$ \\
\hline
kinematics & 13\% & 6\% & 69\% \\
kin. + $E_{iso}<3$~GeV  & 0.7\% & 0.1\% & 64\% \\
kin. + $E_{iso}/p_{T}^{lead}< 0.1$ & 0.7\% & 0.1\% & 66\% \\
\hline\hline
\end{tabular}
\caption{Efficiency rates of various physics objects reconstructed as $\tau_h$, after different requirements: kinematics, absolute ($E_{iso}< 3$~GeV) or relative ($E_{iso}/p_{T}^{lead}< 0.1$) leading track isolation.
Rates are integrated over the full samples.}
\label{tab:misid_rates}
\end{table}

The results, reported in Table~\ref{tab:misid_rates}, indicate misidentification rates 
for light-flavored jets and for b-jets, separately. The $\tau_h$ reconstruction efficiency is also reported.
In the case of electrons, fake $\tau_h$ candidates can only arise when PandoraPFA misidentifies 
an electron as a charged hadron, since \texttt{TauFinder} only clusters the PF objects provided as input. 
This issue is addressed by applying a cut on the 
cluster \emph{electromagnetic fraction} (EMF), which effectively suppresses the misidentification of electrons as pions (Sec.~\ref{sec:emf}). 
In contrast, the fake $\tau$ 
originate primarily from the difficulty in discriminating 
jets from genuine $\tau$ decays.  

The development of robust rejection techniques against fake $\tau_h$ 
implemented in this work is a first approach that can be further improved by dedicated studies,
and it will be a key element of future \texttt{TauFinder} updates. In particular, further analyses of the 
misidentification rates as a function of $p_T$ and $\eta$ are required, and the implementation of a 
multivariate approach, such as a boosted decision tree (BDT), to discriminate jets from real 
$\tau_h$ candidates will be an essential step toward achieving the performance needed for Higgs physics studies at a muon collider.

\subsection{Tau energy corrections}
\label{sec:tau_corrections}

\begin{figure}[H]
\centering
\sidecaption
\includegraphics[width=0.5\textwidth,clip]{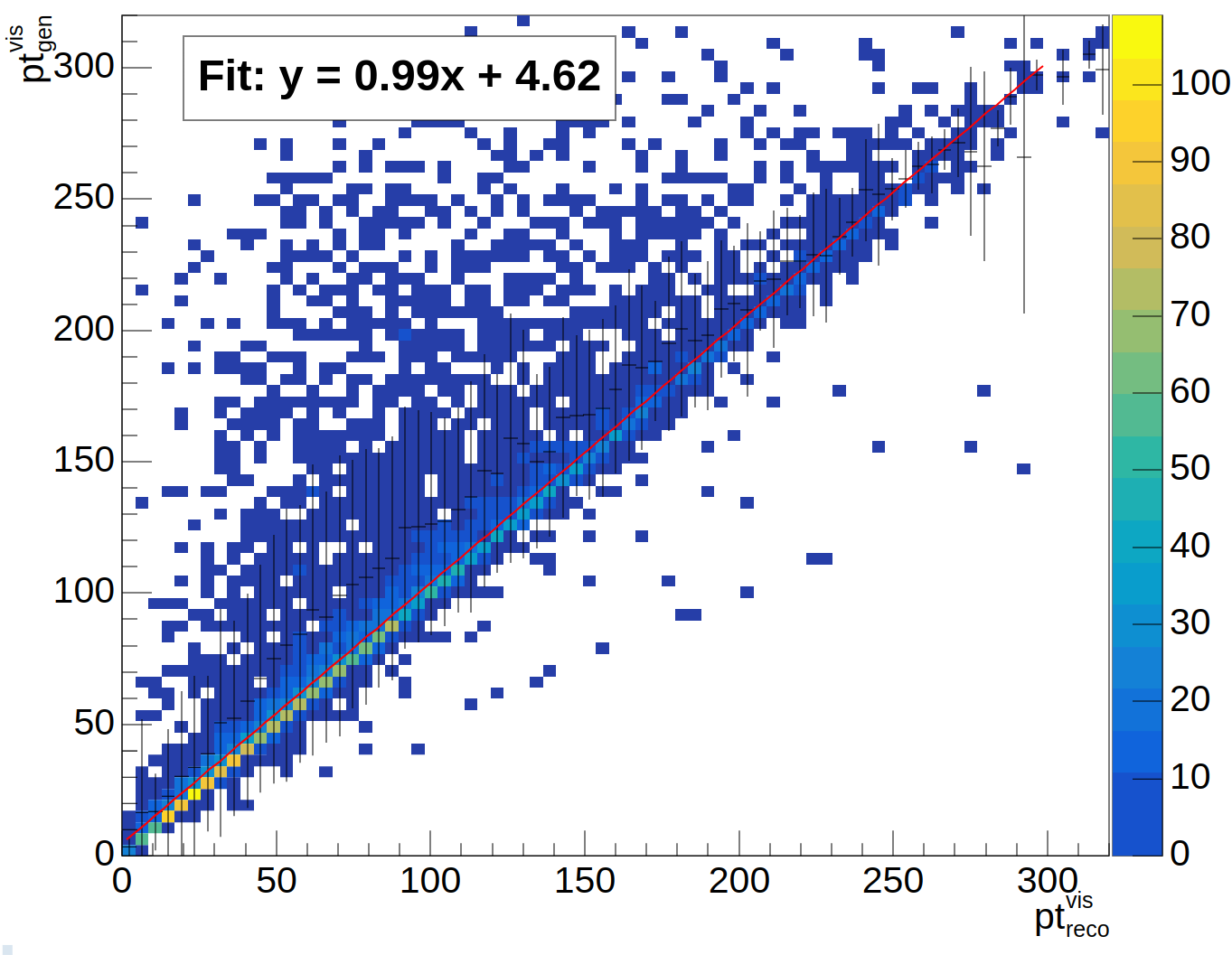}
\caption{Correlation between reconstructed and generator-level transverse momentum
of visible $\tau$ decay products in the $\tau$-gun sample. The linear fit is used to derive the $\tau$ energy correction applied in the main analysis.}
\label{fig:tau_energy_correction}
\end{figure}

When the visible products of the $\tau$ leptons interact with the detector, they are reconstructed with an energy that can differ from their true energy, due to both detector effects and statistical fluctuations. To correct for this bias, a dedicated study was performed using $\tau$ lepton particle-gun samples. In particular, the transverse momentum of the reconstructed visible decay products, $p^{\text{vis}}_{T,\text{reco}}$, was compared to the corresponding generator-level value $p^{\text{vis}}_{T,\text{gen}}$, for each decay mode. The resulting correlation for all hadronic $\tau$ decay modes is shown in Fig.~\ref{fig:tau_energy_correction}, although the energy correction parameters are obtained separately for each decay mode (1P0N, 1P1N, 3P), using dedicated linear fits such as:
\[
p^{\text{vis}}_{T,\text{gen}} = a \cdot p^{\text{vis}}_{T,\text{reco}} + b
\]

This simple linear parametrization may be used to correct the energy of each reconstructed $\tau_h$ in the main analysis. The corrected energy is

\[
E^{\text{corr}} = a \cdot E^{\text{reco}} + b ,
\]

where $E^{\text{reco}}$ is the reconstructed visible energy of the $\tau_h$ candidate. This correction may ensures that the reconstructed kinematical distributions match more closely the true ones, and it can be applied systematically to all $\tau_h$ candidates in the $H \to \tau_h \tau_h$ analysis.
In this study, energy corrections are not applied.

\section{Estimation of the statistical uncertainties on the cross section of the $H\rightarrow\tau^+\tau^-$ process}

The study of the $H \to \tau^+\tau^-$ decay provides direct sensitivity to the Higgs coupling to leptons, offering a powerful test of the SM and a probe for possible New Physics effects. This section focuses on the fully hadronic channel, $H \to \tau_h \tau_h$, reconstructed with the \texttt{TauFinder} algorithm described in Section~\ref{sec:tau_reco_id}. The analysis is performed in the context of a 10~TeV Muon Collider, with the goal of estimating the statistical precision on the cross section measurement. At this stage, the BIB is not yet included in the reconstruction, but preparatory cuts on transverse momentum and detector timing are applied to allow for its future integration.

\subsection{Signal and background samples}

\begin{figure}[H]
\centering
\sidecaption
\includegraphics[width=0.50\textwidth,clip]{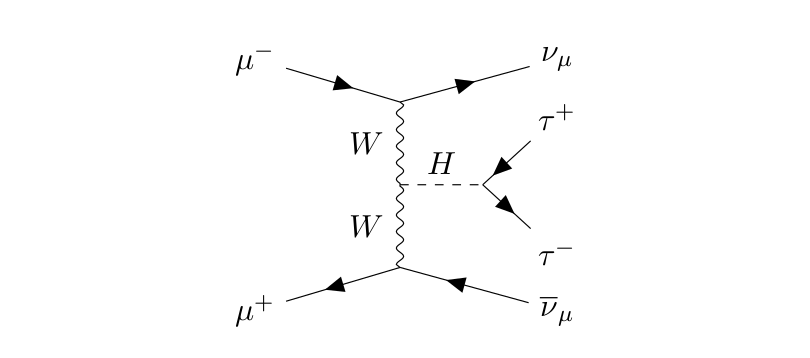}
\caption{Feynman diagram of the $H \to \tau^+ \tau^-$ signal process under investigation.}
\label{signal}       
\end{figure}

The signal under study is the Higgs boson production via $WW$ fusion at a 10~TeV Muon Collider, followed by its decay into a pair of tau leptons, $\mu^+ \mu^- \to H \nu_\mu \bar{\nu}_\mu$, $H \to \tau^+ \tau^-$ (Figure~\ref{signal}). The cross section for this process, as obtained from MadGraph5, is $\sigma = 52.17$~fb. Only the dominant $WW$ fusion contribution is considered, while subleading production mechanisms such as $ZZ$ fusion and Higgs-strahlung are neglected in this first study.

\begin{figure}[H]
\centering
\sidecaption
\includegraphics[width=0.5\textwidth,clip]{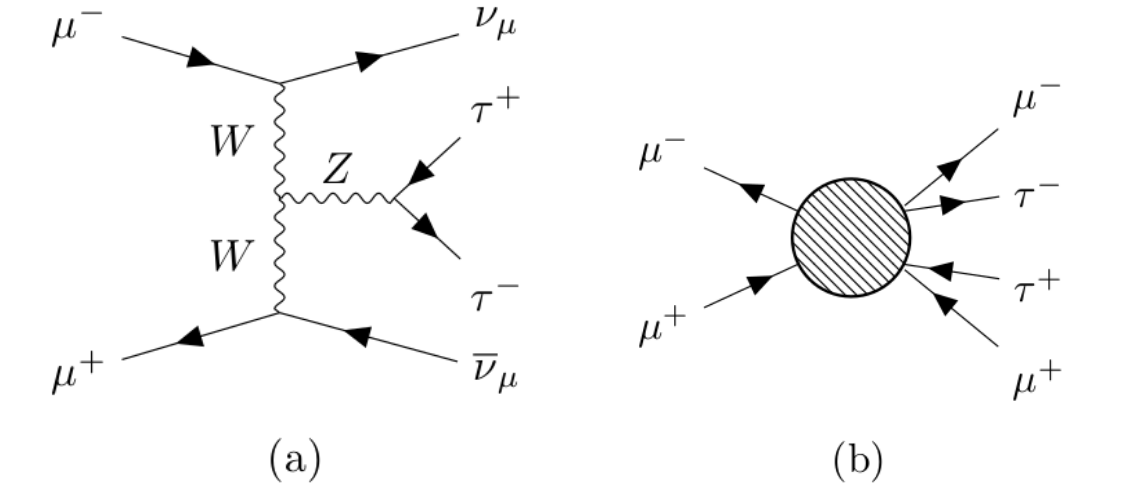}
\caption{Feynman diagrams of the two main background processes in this analysis under investigation: (a) $\mu^+ \mu^- \to Z \nu_\mu \bar{\nu}_\mu$, $Z \to \tau^+ \tau^-$; (b) $\mu^+ \mu^- \to \tau^+ \tau^- \mu^+ \mu^-$.}
\label{bckg}       
\end{figure}

The main irreducible backgrounds are divided into two categories: (i) inclusive $\mu^+ \mu^- \to \tau^+ \tau^- \nu_\mu \bar{\nu}_\mu$ processes, with a total cross section of 127.4~fb, largely dominated by $\mu^+ \mu^- \to Z \nu_\mu \bar{\nu}_\mu$, $Z \to \tau^+ \tau^-$, and (ii) $\mu^+ \mu^- \to \tau^+ \tau^- \mu^+ \mu^-$ processes, with a cross section of 288.6~fb (Figure~\ref{bckg}). In the latter case, most of the final state muons are produced in the forward region outside the detector acceptance, making the visible final state indistinguishable from the signal. In this analysis, all type (ii) processes were conservatively treated as background. 

Backgrounds originating from fake $\tau_h$ candidates were not included at this stage, given the small misidentification rates observed in Sec.~\ref{sec:taufakes}. However, they should be incorporated in future more complete studies.

\subsection{Kinematical variables}

\begin{figure}[H]
\centering
\sidecaption
\includegraphics[width=0.48\textwidth,clip]{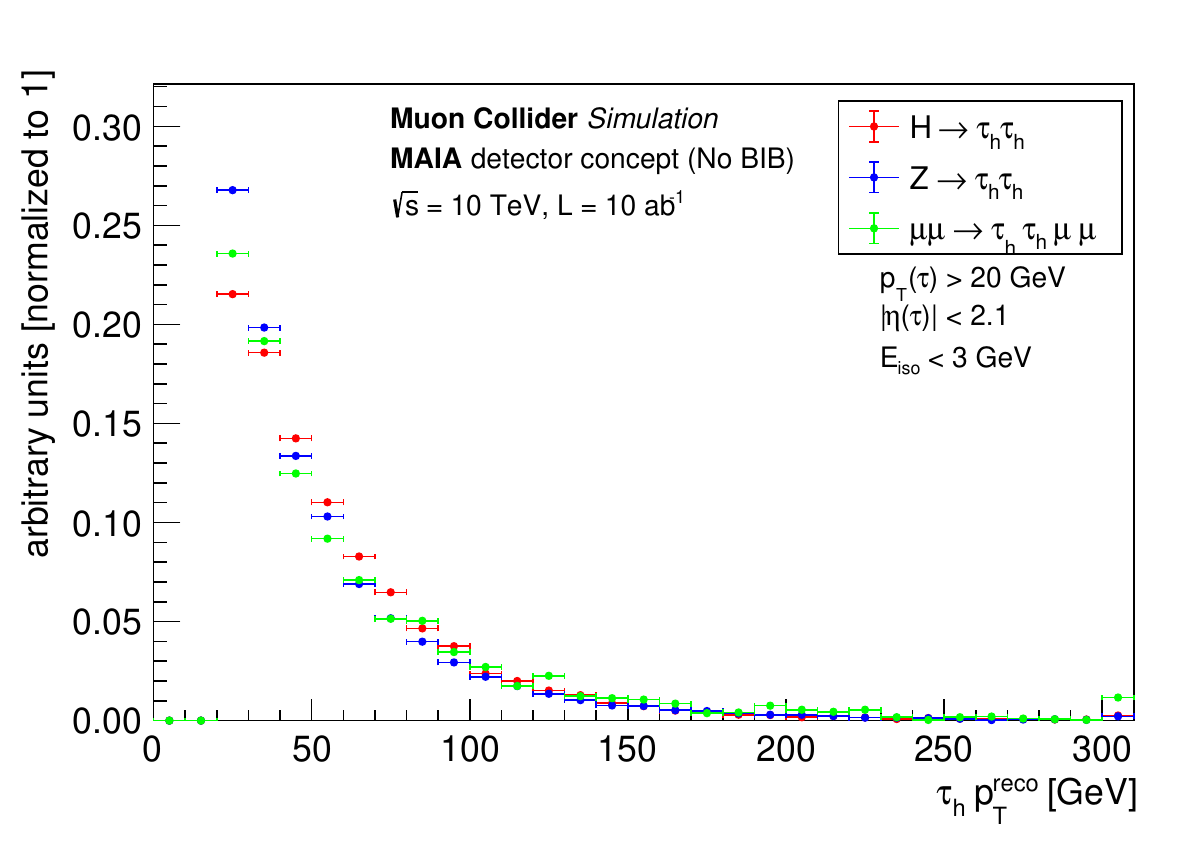}
\caption{Transverse momentum distribution of the reconstructed $\tau_h$ candidates for the signal ($H \to \tau^+ \tau^-$) and background samples ($Z \to \tau^+ \tau^-$ and $\mu^+ \mu^- \to \tau^+ \tau^- \mu^+ \mu^-$).}
\label{fig:pt_tausamples}       
\end{figure}

To characterise the kinematics of the reconstructed $\tau_h$ candidates and to identify possible differences between the signal and background processes, several observables were studied. Figure~\ref{fig:pt_tausamples} shows the $p_T$ distributions of the reconstructed $\tau_h$ candidates for signal and background samples.

\begin{figure}[H]
\centering
\sidecaption
\includegraphics[width=0.48\textwidth,clip]{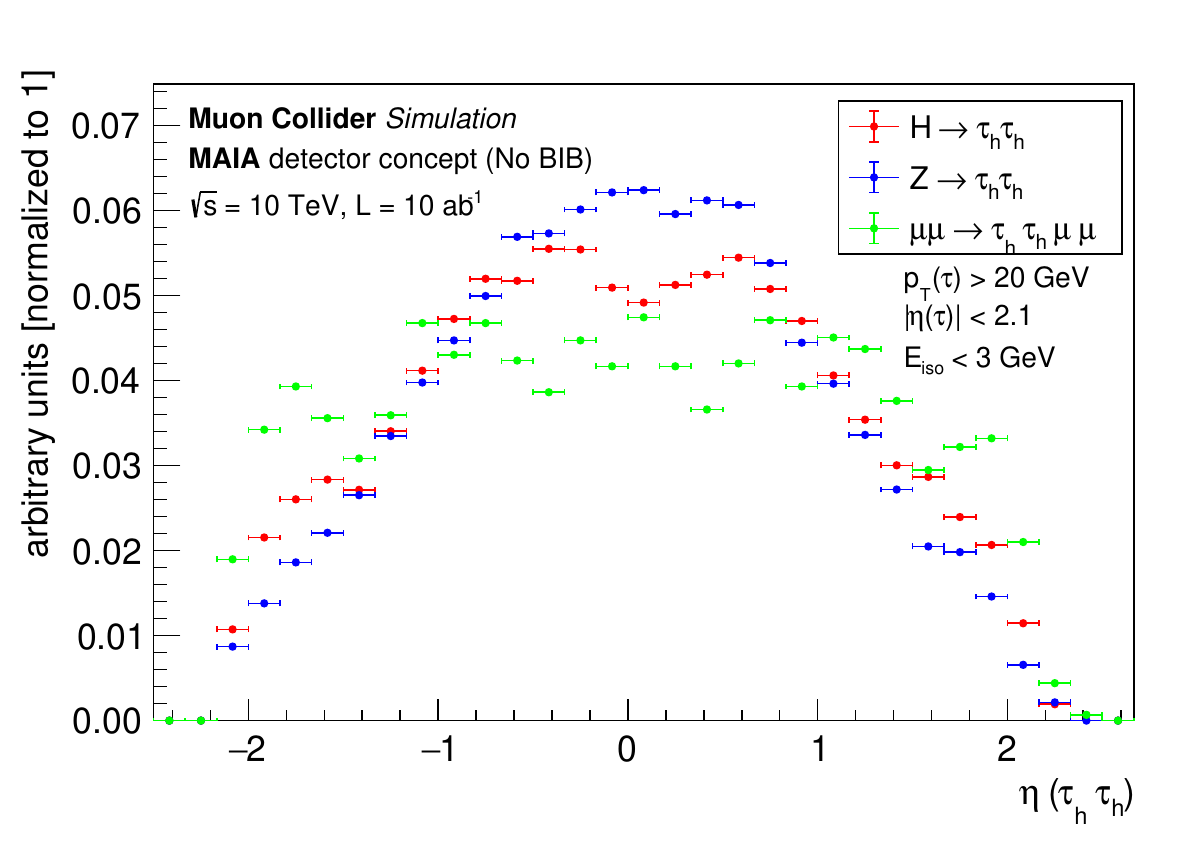}
\caption{Pseudorapidity distribution of the reconstructed $\tau_h$ candidates for the three considered samples. 
}
\label{fig:eta_tauh}       
\end{figure}

The pseudorapidity distributions of the reconstructed $\tau_h$ candidates are presented in Figure~\ref{fig:eta_tauh}.  
Only $\tau_h$ within the detector acceptance, defined by $|\eta| < 2.1$, are considered in the analysis.  
This requirement ensures that the decay products are fully contained within the tracker and calorimeter systems.  

\begin{figure}[H]
\centering
\sidecaption
\includegraphics[width=0.48\textwidth,clip]{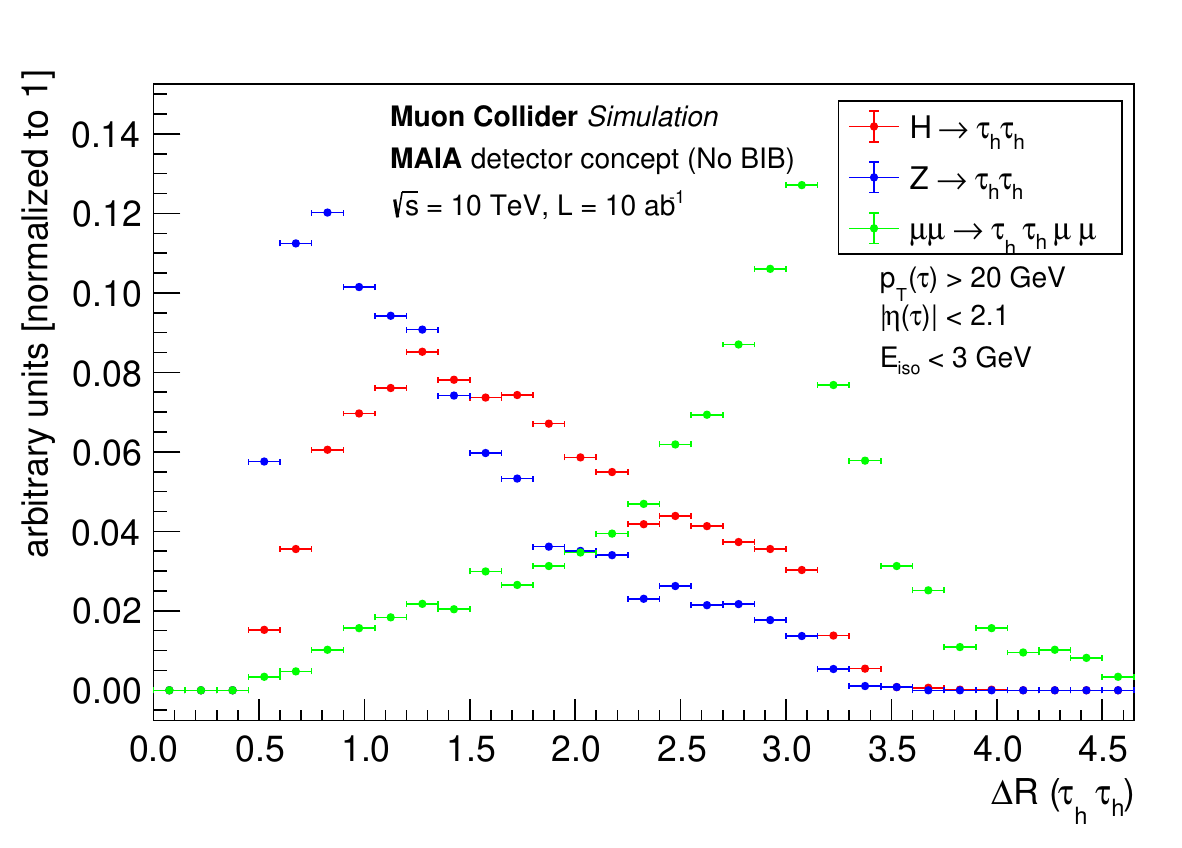}
\caption{Distribution of the angular distance $\Delta R$ between the two reconstructed $\tau_h$ candidates for signal and background samples.}
\label{fig:deltaR_tausamples}       
\end{figure}

In addition, the angular separation $\Delta R$ between the two reconstructed $\tau_h$ candidates
is shown in Figure~\ref{fig:deltaR_tausamples}. 
The distributions are distinctly different and could be used to further separate signal from backgrounds processes.

\subsection{Event selection}

A simple event selection was performed by requiring the presence of exactly two oppositely charged reconstructed $\tau_h$ leptons (with either one or three reconstructed charged hadrons and no requirements on neutral particles). A pseudorapidity requirement of $|\eta|<2.1$ is applied to the reconstructed $\tau_h$ candidates to ensure they fall within the detector acceptance region. A cut of $E_{\text{iso}} < 3$ GeV is applied to select well-isolated $\tau_h$ candidates and suppress background contamination. In order to remove soft backgrounds, also in view of future inclusion of the soft particles from the BIB, a transverse momentum cut $p^{\text{vis}}_{T,\text{reco}} > 20$~GeV is required for the $\tau_h$ candidates. A requirement on the cluster electromagnetic fraction (EMF) was also applied (Sec.~\ref{sec:emf}). The efficiencies of the different samples after the event selection are reported in Table~\ref{tab:selection}.

\begin{table}[h!]
    \centering
    \includegraphics[width=\linewidth]{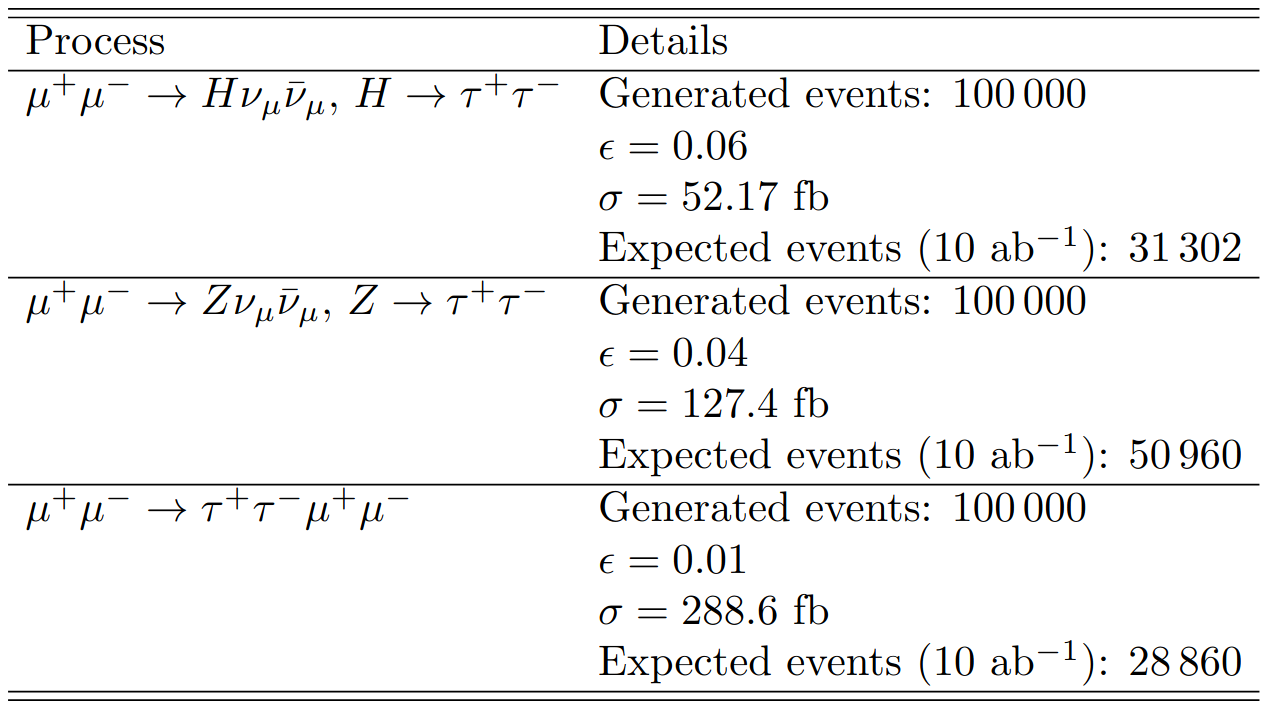} 
    \caption{Number of generated events and those that passed the selection, and the
final efficiency $\epsilon$ for the signal and the two main background processes, for an integrated luminosity of 10~ab$^{-1}$. 
Cross section values are extracted from Madgraph.
}
    \label{tab:selection}
\end{table}

Such low efficiencies, despite the relatively mild selection requirements, can be explained by several factors. First, the shift of the reconstructed transverse momentum distribution towards lower values, due to the undetected neutrinos in the final state, leads to a loss of signal events when applying the $p_T$ cut. The reconstruction efficiency of \texttt{TauFinder}, discussed in Section~\ref{sec:tau_reco_id}, also reduces the number of selected events. 
Finally, only about 42\% (65\% for each of the $\tau_h$ decays) of the signal events correspond to the fully hadronic $\tau^+\tau^-$ final state considered in this study.

\subsection{Fit procedure and results}

\begin{figure}[ht]
\centering
\includegraphics[width=0.48\textwidth]{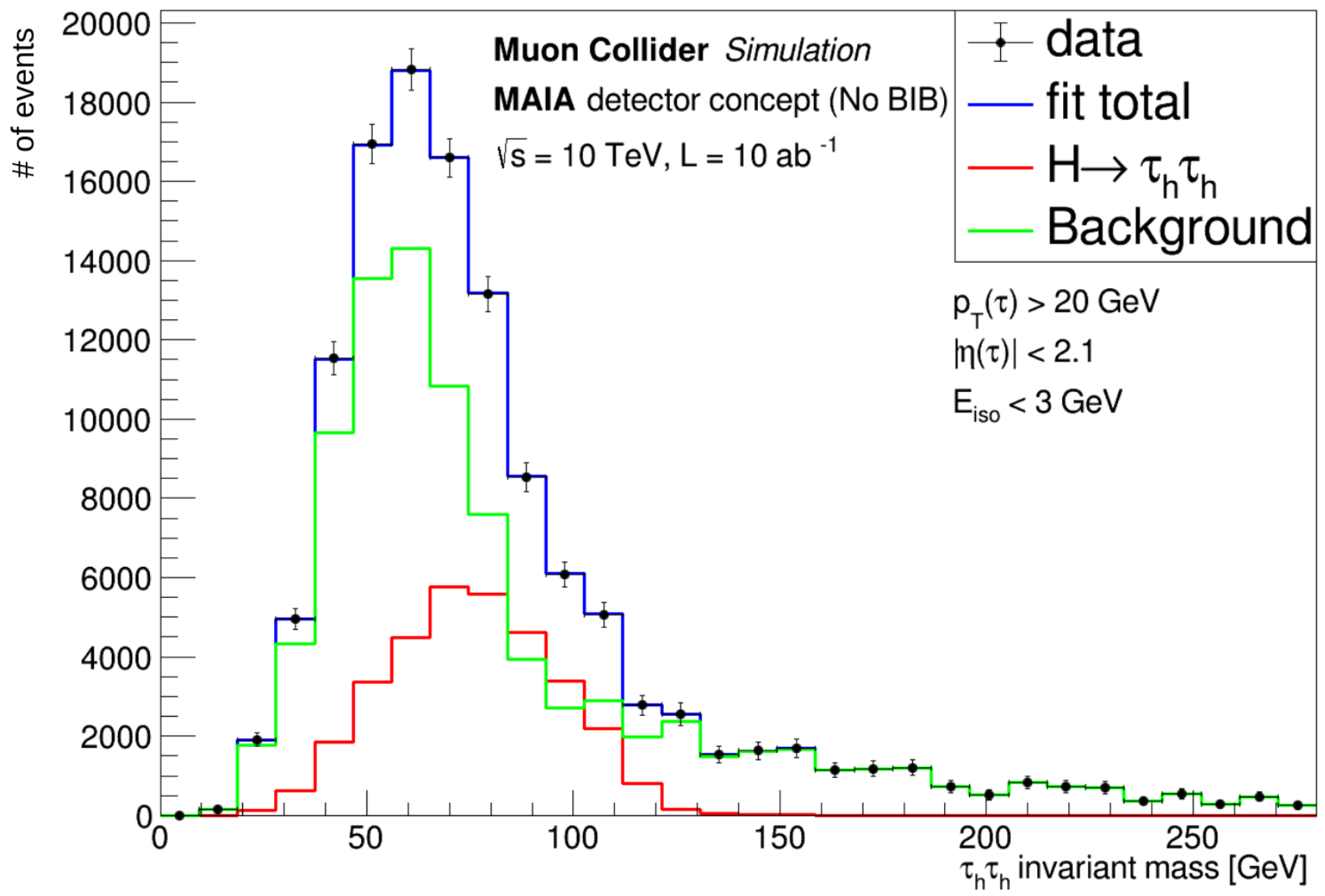}
\caption{Fit of the visible invariant mass distribution of the $\tau_h\tau_h$ system at 
$\sqrt{s}=10~\mathrm{TeV}$ for an integrated luminosity of 10~ab$^{-1}$. The black points represent the pseudo-data, while the red and 
green histograms show the fitted signal and background components, respectively. 
The background includes DY ($\mu\mu\to\nu\bar{\nu}Z$) and $\mu\mu\to\tau\tau\mu\mu$ processes (Table~\ref{tab:selection}).
The blue line is the total fit. The EMF$<$1 and $E_{iso}<$3~GeV selection cuts are also applied.}
\label{fig:mass_fit}
\end{figure}

The signal extraction was performed using the distribution of the visible invariant mass of the $\tau_h\tau_h$ system. The signal and background templates were obtained from the simulated samples after the full event selection and normalized to the expected integrated luminosity of 10~ab$^{-1}$. 
A binned maximum likelihood fit was then performed to the distribution of 
$m_{\tau_h\tau_h}^{\mathrm{vis}}$, as illustrated in Fig.~\ref{fig:mass_fit}. 
In the fit, the signal and background normalizations were left free to float and determined directly 
from the pseudo-data. The fitted signal yield was then converted into a measurement of the production cross section via
\begin{equation}
\sigma(H \to \tau_h \tau_h) = \frac{N_{\mathrm{sig}}}{\epsilon\cdot 
L}
\end{equation}
where $N_{\mathrm{sig}}$ is the number of fitted signal events, $\epsilon$ the total selection 
efficiency, and 
L the integrated luminosity. 

The statistical uncertainty on the cross section was evaluated by generating 
pseudo-experiments in which the bin contents of the invariant mass distribution were 
fluctuated according to Poisson statistics, and the fit procedure was repeated on each. 
The width of the distribution of fitted signal yields provides the estimate of the 
statistical error. 
%
The resulting relative 
statistical uncertainty on the 
cross section of the $H\rightarrow \tau_h\tau_h$ process is
\[
\frac{\Delta\sigma}{\sigma} = 1.3\% .
\]

\section{Discussion}
\label{sec:disc}

A meaningful benchmark against which to compare our result is the projection of Ref.~\cite{refdelphes} obtained with \texttt{Delphes}~\cite{delphes}
fast simulation at $\sqrt{s} = 10$~TeV for an equivalent integrated luminosity. 
The reported statistical precision on the $H \to \tau\tau$ cross section via $W^+W^-$ fusion is $\Delta\sigma/\sigma = 1.1\%$, 
which is comparable to the result obtained in this work.
The agreement indicates that our full simulation analysis yields a sensitivity comparable to the fast-simulation expectation. 
Both results are obtained without the inclusion of the BIB.

\begin{figure}[ht]
\centering
\includegraphics[width=0.48\textwidth]{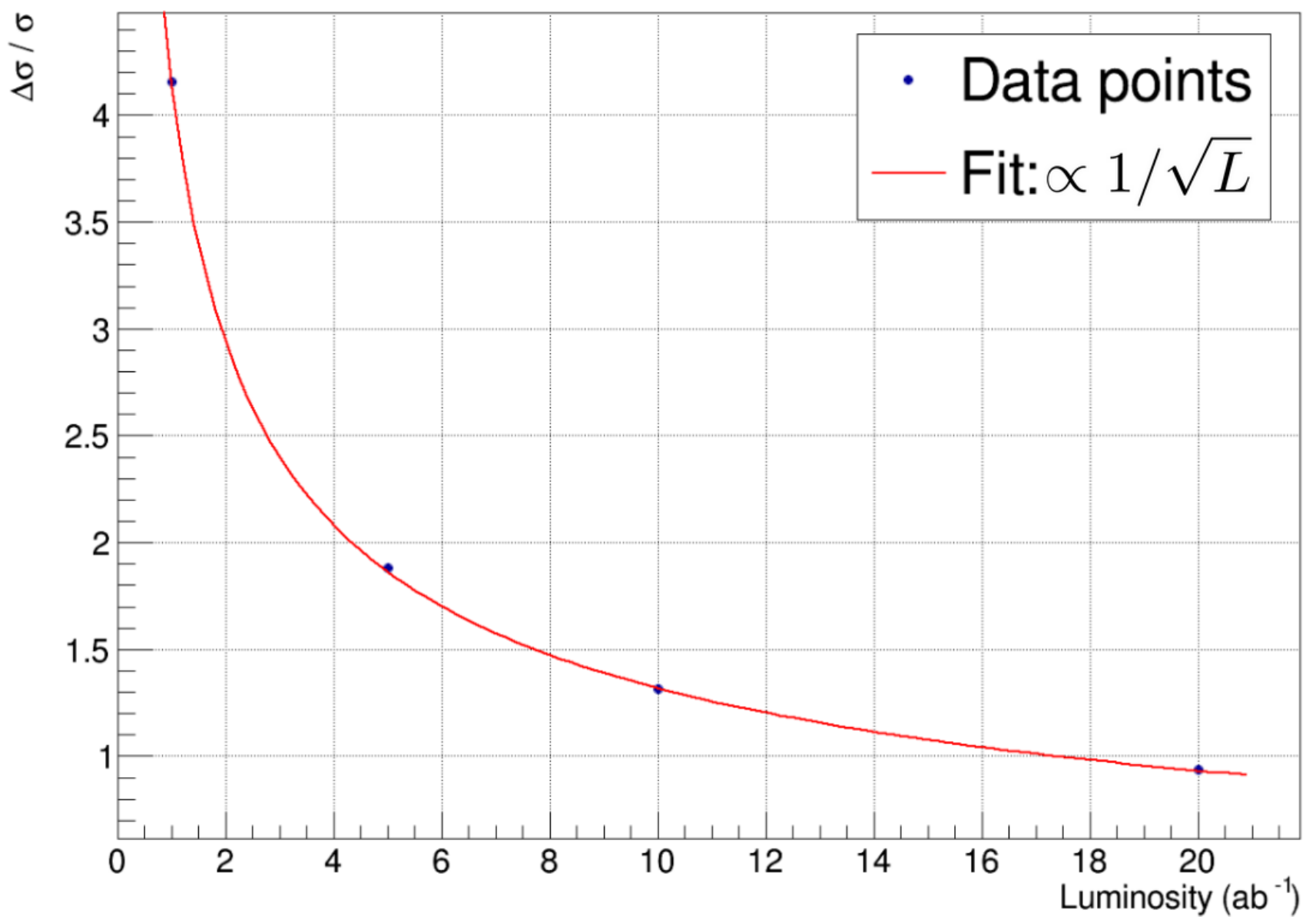}
\caption{Expected relative statistical uncertainty on the $H \rightarrow \tau^+ \tau^-$ cross section as a function of the integrated luminosity. $L = 20~\mathrm{ab^{-1}}$ corresponds to $L = 10~\mathrm{ab^{-1}}$ taken in each of the 2 predicted interaction points of a Muon Collider facility. For $L = 1~\mathrm{ab^{-1}}$, the current result compares to a previous 3~TeV study (5.3\%) \cite{lorenzo}.}
\label{fig:uncertainty_vs_lumi}
\end{figure}

It is worth noting that a previous study at $\sqrt{s} = 3$~TeV with an integrated luminosity of $L = 1~\mathrm{ab^{-1}}$ for the $H \rightarrow \tau^+ \tau^-$ channel at the Muon Collider~\cite{lorenzo} reported a statistical uncertainty of about 5.3\%.
The current analysis at $\sqrt{s} = 10$~TeV yields a relative statistical uncertainty of $\Delta\sigma / \sigma = 4.2\%$ for the same integrated luminosity, as shown in Fig.~\ref{fig:uncertainty_vs_lumi}. 
Results are extrapolated to a total integrated luminosity of 20 $~\mathrm{ab^{-1}}$ corresponding to 10 $~\mathrm{ab^{-1}}$ collected at each of the two anticipated interaction points of the Muon Collider facility.

Larger yields, higher center-of-mass energy, and an improved \texttt{TauFinder} algorithm used in this study help explaining the improved precision. Further improvements in the \texttt{TauFinder} algorithm are foreseen in the near future.

\subsection{Comparison to other Future Colliders}

The current experimental precision on the $\kappa_{\tau}$ parameter, as measured by the CMS \cite{kappatau_cms} and ATLAS \cite{kappatau_atlas} collaborations, is still limited to about 8\%. 
Looking ahead, the comparison can be extended to the projected sensitivities at other future colliders. 
The HL-LHC is expected to reach a precision of about $1.9\%$, while the FCC projections point to an ultimate precision of about $0.44\%$. 
The result presented here, obtained at the $10~\text{TeV}$ Muon Collider, is therefore already competitive with the HL-LHC, and approaches the level of precision expected at the FCC. 
With further improvements in event reconstruction and in the dedicated $H \to \tau\tau$ analysis, the Muon Collider could thus provide a highly competitive measurement of the $\tau$ Yukawa coupling.

\subsection{Future improvements}

Several improvements can be envisaged to enhance the sensitivity of the 
$H \to \tau\tau$ analysis at the Muon Collider.  
A first natural extension is the use of a multivariate technique such as a 
Boosted Decision Tree (BDT) to optimize the discrimination between signal 
and background. While in this work the signal extraction was based solely on the visible invariant mass of the $\tau_h\tau_h$ system, the addition of a BDT trained on angular and kinematic variables could significantly increase the signal-background separation.  

A second improvement concerns the mitigation of 
jet--$\tau_h$ misidentification. 
A dedicated BDT could be trained to reduce the contribution of light- and $b$-jets misidentified as hadronic $\tau$s, thus further 
improving the purity of the selected sample.  

Another important source of improvement is related to the $\tau$ 
reconstruction. In particular, the \texttt{TauFinder} algorithm can be 
optimized to better handle the cases where electrons or charged pions are 
misreconstructed as hadronic $\tau$s. 
In this analysis, this effect was only partially corrected by applying a cut on the electromagnetic fraction (EMF $< 1$), which proved effective in reducing the contamination from electron-like candidates. However, a more robust solution would require a dedicated revision of the algorithm, possibly including new identification variables.

Overall, the implementation of advanced multivariate classifiers, together 
with a dedicated refinement of the $\tau$ reconstruction, has the potential to substantially improve the sensitivity to the $H \to \tau\tau$ process.

\section{Conclusions and Outlook}

The first full-simulation study of the $H \to \tau^+\tau^-$ process at a 10~TeV Muon Collider was presented, focusing on the fully hadronic channel. Using the \texttt{TauFinder} algorithm 
and a fit to the visible invariant mass, we obtained a relative statistical uncertainty on the cross section of 
\[
\frac{\Delta\sigma}{\sigma} = 1.3\% \, ,
\] 
comparable to \texttt{Delphes} projections.
This demonstrates the robustness of the analysis with a realistic detector simulation, and highlights the competitive potential of the Muon Collider compared to HL-LHC (1.9\%) and FCC (0.44\%) sensitivities.

Future improvements could further enhance the precision, in particular the use of multivariate 
techniques (e.g.\ BDTs) to improve signal--background separation, refined $\tau$ reconstruction 
to reduce $\tau_h$ misidentification, and the inclusion of beam-induced backgrounds. With these developments, the uncertainty on the $H \to \tau\tau$ cross section could be reduced well 
below the current result, confirming the Muon Collider as a unique facility for precision Higgs physics at the percent and sub-percent level.

\section*{Acknowledgements}

Special thanks go to the Yale and Madison groups of the US $\tau$ Studies Group 
for their collaboration and insights.  
Finally, I gratefully acknowledge 
the financial support provided by the EU project “MuCol HORIZON-INFRA-2022-DEV-01-01”, 
providing the opportunity and framework to carry out this research.

\bibliography{BibTexFile}
%

\end{document}